\documentclass[runningheads,a4paper]{llncs}

\usepackage{amssymb}
\usepackage{amsmath}
\usepackage{amsfonts}
\usepackage{amsthm}
\usepackage{mathrsfs}
\usepackage{graphicx}
\usepackage{url}
\usepackage{algorithm}
\usepackage{algorithmic}
\newcommand{\keywords}[1]{\par\addvspace\baselineskip
\noindent\keywordname\enspace\ignorespaces#1}

\begin{document}

\mainmatter  

\title{Large-scale Reservoir Simulations on IBM Blue Gene/Q}

\titlerunning{Large-scale reservoir simulations on IBM Blue Gene/Q}

%
%
\author{Hui Liu \and Kun Wang \and Zhangxin Chen}
\authorrunning{Large-scale reservoir simulations on IBM Blue Gene/Q}

\institute{University of Calgary\\
2500 University Dr NW, Calgary, AB, Canada, T2N 1N4 \\
\{hui.j.liu, wang30, zhachen\}@ucalgary.ca
}

\maketitle

\begin{abstract}
This paper presents our work on simulation of large-scale reservoir models on
IBM Blue Gene/Q and studying the scalability of our parallel reservoir simulators.
An in-house black oil simulator has been implemented. It uses MPI for communication and is
capable of simulating reservoir models with hundreds of millions of grid cells.
Benchmarks show that our parallel simulator are thousands of times faster than
sequential simulators that designed for workstations and personal computers,
and the simulator has excellent scalability.

\end{abstract}

\section{Introduction}
Nowadays, large-scale reservoir simulations are becoming more and more popular
in the oil and gas industry in order to simulate complex geological
models. However, when a model is large enough, a simulator may take days or even
weeks to finish one run using regular workstations and personal computers.
This problem can also be observed in black oil, compositional and thermal simulations.
Efficient computational methods and fast reservoir simulators
should be investigated.

Reservoir simulations have been studied for decades and various models and methods have been developed.
Coats studied black oil, compositional and thermal models, and he also investigated
numerical methods, linear solver, preconditioner, grid effects and stability
issues in his publications \cite{c1,c2,c3,c4,c5,c6}.
Kaarstad et al. \cite{PS-Kaa} implemented a parallel two-dimensional two-phase oil-water simulator,
which could solve problems with millions of grid cells.
Rutledge et al. \cite{PS-Rut} implemented a compositional simulator
for parallel computers using the IMPES (implicit pressure-explicit saturation) method.
Shiralkar et al. \cite{PS-Shi} developed a portable parallel production qualified simulator,
which could run on a variety of parallel systems.
Killough et al. \cite{PS-Kil2} studied locally refined grids in their parallel simulator.
Dogru et al. \cite{PS-Dogru2} developed a parallel black oil simulator,
which was highly efficient and was capable of simulating models with up to one billion cells.
Zhang et al. developed a scalable general-purpose platform to
support adaptive finite element and adaptive finite
volume methods, which was also applied to reservoir simulations using Discontineous Galerkin method \cite{phg,phg-quad,kwang}.
For many reservoir simulations, most of the running time is spent on the solution of linear systems.
We know that the most important is to develop efficient preconditioners.
Many preconditioners have been proposed, such as constrained pressure residual (CPR) methods \cite{CPR-old,CPR-cao},
multi-stage methods \cite{Study-Two-Stage}, multiple level preconditioners
\cite{mlp} and fast auxiliary space preconditioners (FASP) \cite{FASP,FASP2}.
Chen et al. studied parallel reservoir simulations and developed a family of
CPR-like preconditioners for black oil simulations and compositional simulations,
including CPR-FP, CPR-FPF and CPR-FFPF methods \cite{bos-pc}.

A black oil simulator has been developed based on our in-house parallel platform.
The black oil model has three mass conservation equations for three components (water, gas and oil).
The system is fully coupled nonlinear system, which is solved
by inexact Newton-Raphson methods, and structured grids and finite difference methods are applied.
The performance of the black oil simulator is studied on IBM Blue Gene/Q system using large-scale
reservoir models for standard black oil model and two-phase oil-water model.
Numerical experiments show that our simulator is scalable
and it is capable of simulating models with hundreds of millions of grid cells.

\section{Reservoir Simulation Models}
\label{model}

The black oil model and its simplified model, two-phase oil-water model,
are briefly introduced here.

\subsection{Black Oil Model}

The black oil model has three phases (water, oil and gas), and three components.
The model assumes that there is no mass transfer between water phase and the other two phases,
and gas component can exist in gas and oil phases.
Oil component is also assumed that it can exist in oil phase only. The reservoir is isothermal
and no energy change is considered.

The Darcy's law is applied for black oil model, which establishes a relationship between volumetric flow rates of
three components and their pressure changes in a reservoir, which is described as:
\begin{equation}
Q = - \frac{\kappa A\Delta p}{\mu L},
\end{equation}
where $\kappa$ is the absolute permeability of rock, $A$ is a cross-section area, $\Delta p$ is the pressure difference, $\mu$
is viscosity of fluid, and $L$ is the length of a porous medium.
In three-dimensional space, the differential form of Darcy's law is:
\begin{equation}
q = \frac{Q}{A} = - \frac{\kappa}{\mu} \nabla p.
\end{equation}

By combining Darcy's law, black oil model has the following mass conservation equations for each component:
\begin{equation}
 \left\{
 \begin{aligned}
& \frac{\partial}{\partial t} (\phi s_o \rho_o^o)& = &\nabla \cdot ( \frac{K K_{ro}}{\mu_o} \rho_o^o \nabla \Phi_o) + q_o, \\
& \frac{\partial}{\partial t} (\phi s_w \rho_w)& = &\nabla \cdot ( \frac{K K_{rw}}{\mu_w} \rho_w \nabla \Phi_w) + q_w, \\
& \frac{\partial(\phi \rho_o^g s_o + \phi \rho_g s_g)}{\partial t}& = &
\nabla \cdot ( \frac{K K_{ro}}{\mu_o} \rho_o^g \nabla \Phi_o) + \nabla \cdot ( \frac{K K_{rg}}{\mu_g} \rho_g \nabla \Phi_g) + q_o^g + q_g ,
\end{aligned}
 \right.
\end{equation}
where, for phase $\alpha$ $(\alpha = o,w,g)$,
    $\Phi_{\alpha}$ is its potential, $\phi$ and $K$ are porosity and permeability of a resevoir, and
      $s_{\alpha}$, $\mu_{\alpha}$, $p_{\alpha}$,
    $\rho_{\alpha}$, $K_{r\alpha}$ and $q_{\alpha}$ are
    its saturation, phase viscosity, phase pressure,
    density, relative permeability and production (injection) rate, respectively.
$\rho_o^o$ and $\rho_o^g$ are density of the oil component in the oil phase and
    the density of the solution gas in the oil phase, respectively.
They have the following relations:
\begin{equation}
\label{bos-cs}
 \left\{
 \begin{aligned}
& \Phi_{\alpha} = p_{\alpha} + \rho_{\alpha} g z, \\
& S_o + S_w + S_g = 1,        \\
& p_w = p_o - p_{cow} (S_w),  \\
& p_g = p_o + p_{cog} (S_g),
 \end{aligned}
 \right.
\end{equation}
where $z$ is reservoir depth, $p_{cow}$ is capillary pressure between
water phase and oil phase, $p_\alpha$ is pressure of phase $\alpha$,
and $p_{cog}$ is capillary between gas phase and oil phase.

The properties of fluids and rock are functions of
pressure and saturation.
The pressures of water and gas phases are functions of oil phase pressure and
saturation; see equation \eqref{bos-cs}.
The density of water is a function of its pressure:
\[
\rho_w = \rho_w(p_w) = \rho_w(p_o, s_w),
\]
and the density of the oil phase is a function of its phase pressure and the bubble point pressure:
\[
\rho_o^o = \rho_o^o(p_o, p_b),
\]
where $p_b$ is bubble point pressure. The bubble point pressure is the pressure
at which infinitesimal gas appears. The water viscosity $\mu_w$ is assumed to
be a constant. The oil phase
viscosity is a function of its pressure $p_o$ and the bubble point pressure $p_b$:
\[
\mu_o = \mu_o(p_o, p_b).
\]
The relative permeabilities $K_{rw}$, $K_{ro}$ and $K_{rg}$ are functions of water and gas saturations $S_w$ and $S_g$:
\[
\left\{
\begin{aligned}
& K_{rw} = K_{rw}(S_w), \\
& K_{rg} = K_{rg}(S_g), \\
& K_{ro} = K_{ro}(S_w, S_g),
\end{aligned}
\right.
\]
where $K_{ro}$ is calculated using the Stone II formula.

For real simulations, the relative permeabilities are given by tables or analytic formulas.
Other properties, such as density, viscosity and capillary pressure, have analytic formulas
and they can be calculated by table input too.
With proper boundary conditions and initial conditions, a close
    system is given. Here no flow boundary condition is assumed.

\subsection{Two-Phase Flow Model}

This model is a simplified model of the standard black oil model,
which assumes that the reservoir has two phases, oil and water, and they are immiscible.
The model is similar to black oil model, which is written as \cite{Book-Chen}:
\begin{equation}
\label{eq-two}
 \left\{
 \begin{aligned}
& \frac{\partial}{\partial t} (\phi s_o \rho_o) & = & \nabla \cdot ( \frac{K K_{ro}}{\mu_o} \rho_o \nabla \Phi_o) + q_o \\
& \frac{\partial}{\partial t} (\phi s_w \rho_w) & = & \nabla \cdot ( \frac{K K_{rw}}{\mu_w} \rho_w \nabla \Phi_w) + q_w.
\end{aligned}
 \right.
\end{equation}

\subsection{Well Modeling}

Different well constraints can be set for each active well.
One commonly-used method is a sink-source model. For each perforation block $m$,
its well rate (production or injection) $q_{\alpha,m}$ is calculated by:
\begin{equation}
q_{\alpha,m} = W_i \frac{\rho_\alpha K_{r\alpha}}{\mu_\alpha} (p_{h} - p_\alpha - \rho_\alpha \wp (z_{h} - z)),
\end{equation}
where $p_{h}$ is bottom hole pressure of a well, $W_i$ is its well index,
$z_{h}$ is reference depth of bottom hole pressure, $z$ is depth of the perforated grid block $m$,
and $p_\alpha$ is phase pressure of the perforated grid block, such as oil, gas and water.
$W_i$ can be calculated by several different models. In our simulator, a Peaceman model \cite{PWM}
is chosen.

Many operation constraints and their combinations may be applied to each well at different time stages,
such as a fixed bottom hole pressure constraint, a fixed oil rate constraint, a fixed water
rate constraint, a fixed liquid rate constraint and a fixed gas rate constraint.
When the fixed bottom hole pressure condition is applied to some well,
its bottom hole pressure, $p_{h}$, is known and its well rate $q_{\alpha, m}$ is known
if we have phase pressure of the perforated block.
The constraint equation for the well is
\begin{equation}
p_h = c,
\end{equation}
where $c$ is a constant set by the user input. No known exists for this constraint.

When a fixed rate constraint is applied to a well, its
bottom hole pressure is an unknown. For the fixed water rate constraint,
the equation is
\begin{equation}
\sum_{m} {q_{w,m}} = q_w,
\end{equation}
where $q_w$ is constant.
For the fixed oil rate constraint, its equation is
\begin{equation}
\sum_{m} {q_{o,m}} = q_o,
\end{equation}
where $q_o$ is constant and known.
A well may be applied different constraints at different time period. A schedule can
be set by input, in which users can set operation changes for each well.

\subsection{Numerical Methods}

In this paper, conservative finite difference schemes are employed to discretize these models.
The inexact Newton method is employed to solve the nonlinear equations.
The time term is discretized by the backward Euler difference scheme.
If we let $f^n$ represent the value of a function $f$ at any time step $n$, then its derivative at time step
$(n+1)$ is approximated by
\begin{equation}
(\frac{\partial f}{\partial t})^{n+1} = \frac{f^{n+1} - f^{n}}{\Delta t}.
\end{equation}

The space terms are discretized by cell-centered finite difference method \cite{Book-Chen}.
Here if we assume $d$ is a space direction and $A$ is the area of the corresponding face of a grid cell,
the transmissibility term $T_{\alpha,d}$ can be written as:
\begin{equation}
T_{\alpha,d} = \frac{K K_{r\alpha}}{\mu_{\alpha}} \rho_{\alpha} \frac{A}{\Delta d}.
\end{equation}

\subsubsection{Inexact Newton Method}
The nonlinear system can be represented by
\begin{equation}
\label{bos-nonlinear}
F(x) = 0,
\end{equation}
where $x$ is unknown vector, including oil phase pressure, water saturation and well bottom hole pressure.
For black oil model, a gas saturation (or bubble point pressure) is also included. After linearization,
a linear system, $A x = b$, is obtained in each Newton iteration, where $A$ is Jacobian matrix,
$x$ is unknown to be determined, and $b$ is right-hand side. The standard Newton
method solves the linear system accurately. However, it is computationally expensive and
it is not necessary sometimes. In our implementation, the inexact Newton method is applied,
whose algorithm is described in Algorithm \ref{inewton-alg}.

\begin{algorithm}
\caption{The inexact Newton Method}
\label{inewton-alg}
\begin{algorithmic}[1]
\STATE Give an initial guess $x^0$ and stopping criterion $\epsilon$, let $l = 0$,
 and assemble right-hand side $b$.
\WHILE{$\left\|b \right\| \ge \epsilon$}
\STATE Assemble the Jacobian matrix $A$.
\STATE Find $\theta_l$ and $\delta x$ such that
    \begin{equation}
    \label{inexact-newton-alg}
    \left\| b - A \delta x \right\| \leq \theta_l \left\| b \right\|,
    \end{equation}
\STATE    Let $l = l+1$ and $x = x + \delta x$.
\ENDWHILE
\STATE $x$ is the solution of the nonlinear system.
\end{algorithmic}
\end{algorithm}

The only difference between standard Newton method and inexact Newton method is how to
choose $\theta_l$. Usually the parameter, $\theta_l$, for standard Newton method is fixed and small, such
as $10^{-5}$. The parameter, $\theta_l$, for inexact Newton method is automatically adjusted.
Three different choices are listed as follows \cite{kwang}:
\begin{equation}
\theta_l =
\left\{
\begin{aligned}
\label{choice-inm}
& \frac{\left\| b^l - r^{l-1} \right\|}{\left\| b^{l-1} \right\|}, \\
& \frac{\left\| b^l \right\| - \left\| r^{l-1} \right\|}{\left\| b^{l-1} \right\|}, \\
& \gamma \left( \frac{\left\| b^l \right\|}{\left\| b^{l-1} \right\|} \right)^{\beta},
\end{aligned}
\right.
\end{equation}
where $r^l$ and $b^l$ are residual and right-hand side of $l$-th iteration, respectively. The residual is
defined as,
\begin{equation}
r^l = b^l - A \delta x.
\end{equation}

\subsubsection{Linear Solver}

If a proper matrix ordering (numbering of unknowns) is applied,
the matrix $A$ derived from each Newton iteration can be written as

\begin{equation}
\label{mat-ab}
A = \left(
        \begin{array}{lll}
        A_{pp} \hspace{0.1cm}   & A_{ps} \hspace{0.1cm} & A_{pw}   \\
        A_{sp} \hspace{0.1cm}   & A_{ss} \hspace{0.1cm} & A_{sw}   \\
        A_{wp} \hspace{0.1cm}   & A_{ws} \hspace{0.1cm} & A_{ww}   \\
        \end{array}
        \right),
\end{equation}
where $A_{pp}$ is the matrix corresponding to oil phase pressure unknowns, $A_{ss}$ is
the matrix corresponding to other unknowns in each grid cell, such as water saturation,
gas saturation and bubble point pressure, and $A_{ww}$ is the matrix coefficients
corresponding to well bottom hole pressure unknowns, and other matrices are coupled items.

The matrix $A$ is hard to solve in large-scale reservoir simulations.
Many multi-stage preconditioners have been developed to overcome this problem, such as
CPR, FASP, CPR-FP and CPR-FPF methods.
The key idea is to solve a sub-problem ($A_{pp}$) using algebraic multi-grid methods (AMG).
In this paper, the CPR-FPF method developed by Chen et al. \cite{bos-pc} is applied.
Matrix decoupling techniques are also employed, such as ABF decoupling and Quasi-IMPES decoupling.

\section{Numerical Experiments}
\label{sec-exp}

An Blue Gene/Q from IBM is employed to run reservoir simulations.
The system, Wat2Q, is located in the IBM Thomas J. Watson Research Center.
Each node has 32 computer cards (64-bit PowerPC A2 processor), which has 17 cores.
One of them is for the operation system and the other
16 cores for computation. The system has 32,768 CPU cores for computation.
The performance of each core is really low
compared with Intel processors. However, the system has strong network relative to CPU performance,
and the system is scalable.

\subsection{Oil-water Model}

The SPE10 model is described on a regular Cartesian grid, whose dimensions are
$1,200 \times 2,200 \times 170 $ (ft) \cite{SPE10}.
The model has $60 \times 220 \times 85$ cells ($1.122 \times 10^6$ cells).
It has one injection well and four production wells.
The original model is designed for two-phase oil-water model and it
has around 2.244 millions of unknowns.

\begin{figure}[!htb]
\begin{center}
    \includegraphics[width=0.6\linewidth]{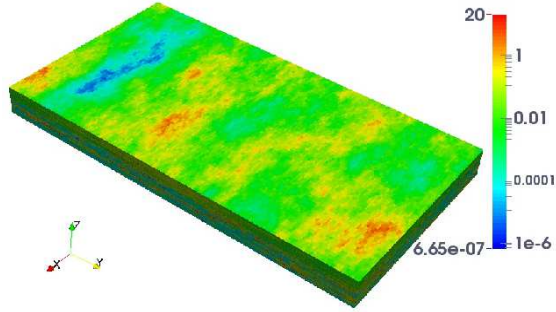}
\end{center}
\caption{Permeability in X Direction of the SPE10 benchmark}
\label{fig-spe10-perm}
\end{figure}

\begin{figure}[!htb]
\begin{center}
    \includegraphics[width=0.6\linewidth]{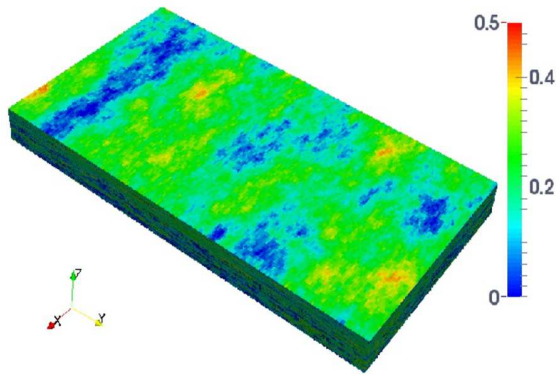}
\end{center}
\caption{Porosity of the SPE10 benchmark}
\label{fig-spe10-poro}
\end{figure}

The model is highly heterogeneous. Its permeability
is ranged from 6.65e-7 Darcy to 20 Darcy, and the x-direction permeability, $K_x$,
is shown in Figure \ref{fig-spe10-perm}.
Its porosity shown in Figure \ref{fig-spe10-poro},
ranges from 0 to 0.5. Data sets for porosity and permeability can be downloaded from SPE10's
official website.
{The relative permeability of water phase is calculated by
\begin{equation}
K_{rw} (s_w) = \frac{(s_w - s_{wc})^2}{(1 - s_{wc} - s_{or})^2},
\end{equation}
and the relative permeability of oil phase is calculated by
\begin{equation}
K_{ro} (s_w) = \frac{(1 - s_{or} - s_w)^2}{(1 - s_{wc} - s_{or})^2},
\end{equation}
where $s_{wc} = s_{or} = 0.2$.
Capillary pressure is ignored.}

\subsection{Numerical Examples}

\begin{figure}[!htb]
\caption{Scalability of {preconditioners}, Example \ref{ex2}}
\begin{center}
    \includegraphics[width=0.7\linewidth, angle=270]{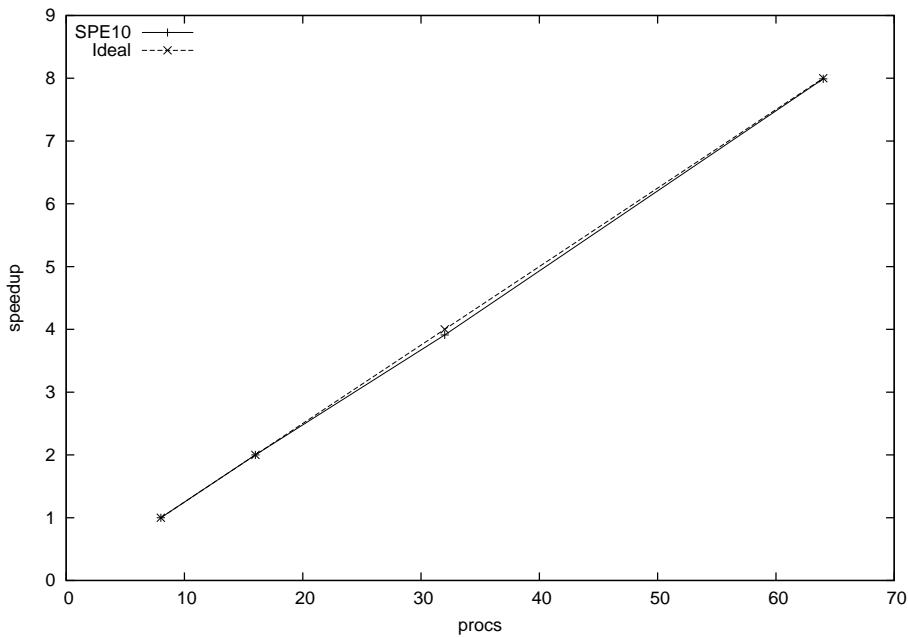}
\end{center}
\label{fig-ex2-cpr}
\end{figure}

\begin{example}
\label{ex2}
The original SPE10 project is simulated. The termination tolerance
for inexact Newton method is $10^{-2}$ and its maximal Newton iterations are 20.
The linear solver BiCGSTAB is applied and its maximal inner iterations are 50.
The Quasi-IMPES decoupling strategy is used.
Simulation period is 2,000 days and maximal time step is 100 days.
Summaries of numerical results are shown in Table \ref{tab-ex2-4} \cite{bos-pc},
and its scalability is shown by Figure \ref{fig-ex2-cpr}.
\end{example}

\begin{table}[!htb]
\centering
\caption{Summaries of Example \ref{ex2}}
\begin{tabular}{lllllll} \\ \hline
  \# Procs  & \# Steps & \# Newton & \# Solver & \# Avg. solver  & Time (s) & Avg. time (s)\\ \hline
   8        & 50 & 298 & 7189 & 24.1 & 27525.6 & 92.3 \\
   16       & 50 & 297 & 7408 & 24.9 & 13791.8 & 46.4 \\
   32       & 51 & 322 & 7467 & 23.1 & 7044.1 & 21.8 \\
   64       & 50 & 294 & 7609 & 25.8 & 3445.8 & 11.7 \\
\hline
\end{tabular}
\label{tab-ex2-4}
\end{table}

Table \ref{tab-ex2-4} presents results for time steps, Newton iterations, total
linear iterations, average linear iterations per Newton iteration, overall running time
and average running time per Newton iteration. From this table, we can see each case has
similar time steps and Newton iterations. The linear solver and preconditioner are robust, where
each Newton iteration terminate in around 25 linear iterations. The table and Figure \ref{fig-ex2-cpr}
show our simulator has excellent scalability.

\begin{figure}[!htb]
    \centering
    \includegraphics[width=0.6\linewidth, angle=270]{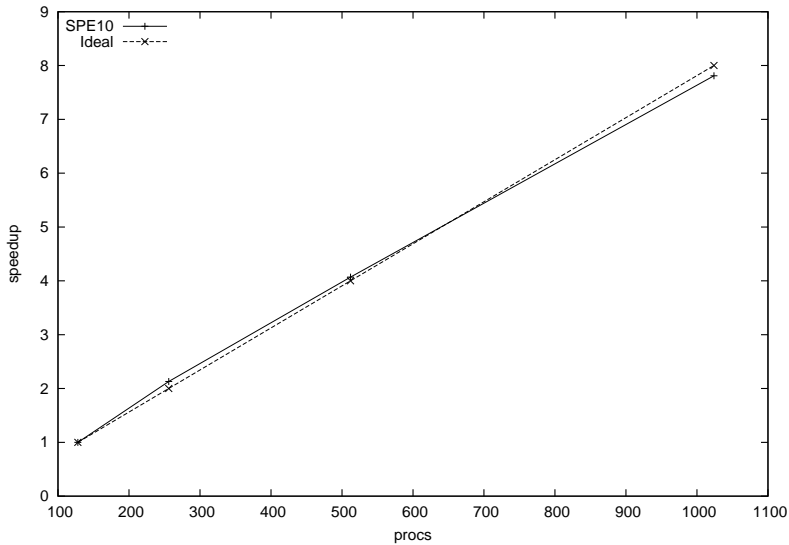}
    \caption{Scalability of Example \ref{ex-scal-ex4}}
    \label{fig-scal-e4}
\end{figure}

\begin{example}
\label{ex-scal-ex4}
This example tests a refined SPE10 case, and each grid cell
is refined into 27 grid cells. It has around 30 millions of grid cells and around 60 millions of unknowns.
The stopping criterion for the inexact Newton method is 1e-3 and its maximal Newton iterations are 20.
The BiCGSTAB solver is applied and its maximal iterations are 100.
Potential reordering and Quasi-IMPES decoupling strategy are applied.
The simulation period is 10 days. Up to 128 computer cards are used.
The numerical summaries are shown
in Table \ref{ex-scal-e4}, and its speedup (scalability) is shown in
Figure \ref{fig-scal-e4}.
\end{example}

\begin{table}[!htb]
\centering
  \caption{Numerical summaries of Example \ref{ex-scal-ex4}}
\begin{tabular}{lllllll} \\ \hline
  \# Procs   & \# Steps & \# Newton & \# Solver & \# Avg. solver  & Time (s) & Avg. time (s)\\ \hline
    128 & 40(1) & 295 & 2470 & 8.3 & 43591.8 & 147.7 \\
    256 & 39 & 269 & 2386 & 8.8 & 20478.4 & 76.1 \\
    512 & 40 & 260 & 2664 & 10.2 & 10709.8 & 41.1 \\
    1024 &39 & 259 & 2665 & 10.2 & 5578.7 & 21.5 \\
\hline
\end{tabular}
  \label{ex-scal-e4}
\end{table}

The numerical summaries in Table \ref{ex-scal-e4}
show the inexact Newton method is robust, where around 40 time steps and around 260 Newton iterations are
used for each simulation with different MPI tasks except the case with 128 MPI tasks due to
one time step cut that contributes 20 Newton iterations.
The linear solver BiCGSTAB and the preconditioner
show good convergence, where the average number of linear iterations for each nonlinear iteration is between 8 and 11.
The results mean our linear solver and preconditioner are effective and robust.
The overall running time and average time for each Newton iteration show our simulator has excellent scalability,
which is almost ideal. The scalability is also
demonstrated by Figure \ref{fig-scal-e4}.

\begin{figure}[!htb]
    \centering
    \includegraphics[width=0.6\linewidth, angle=270]{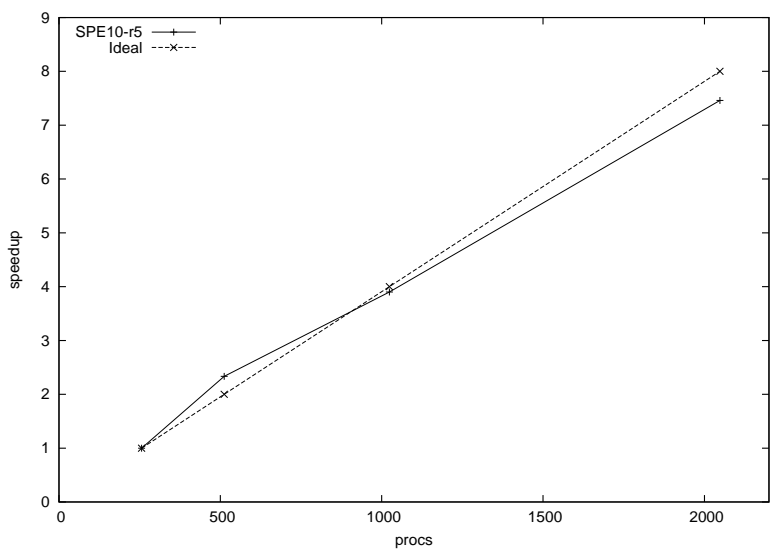}
    \caption{Scalability of Example \ref{ex-scal-ex5}}
    \label{fig-scal-e5}
\end{figure}

\begin{example}
\label{ex-scal-ex5}
This example also tests a refined SPE10 case, where each grid cell
is refined into 125 grid cells. It has around 140 millions of grid cells and around 280 millions of unknowns.
The stopping criterion for inexact Newton method is 1e-2 and its maximal Newton iterations are 20.
The BiCGSTAB solver is applied and its maximal iterations are 100.
The Quasi-IMPES decoupling strategy is applied. The simulation period is 10 days.
The numerical summaries are shown in Table \ref{ex-scal-e5}, and the speedup (scalability) curve is shown in
Figure \ref{fig-scal-e5}.
\end{example}

\begin{table}[!htb]
\centering
  \caption{Numerical summaries of Example \ref{ex-scal-ex5}}
\begin{tabular}{lllllll} \\ \hline
  \# Procs   & \# Steps & \# Newton & \# Solver & \# Avg. solver  & Time (s) & Avg. time (s)\\ \hline
  256  & 57 & 328 & 2942 & 8.9 & 168619.6  & 514.0 \\
  512  & 60 & 328 & 2236 & 6.8 & 72232.4   & 220.2 \\
  1024 & 62 & 341 & 3194 & 9.3 & 43206.5   & 126.7 \\
  2048 & 59 & 327 & 3123 & 9.5 & 22588.8   & 69.0 \\ \hline
\end{tabular}
  \label{ex-scal-e5}
\end{table}

This case is difficult. However, results from Table \ref{ex-scal-e5} show our nonlinear and linear methods
are robust. Each Newton iteration terminate in less than 10 linear iterations. Running time and Figure \ref{fig-scal-e5}
show our simulator has good scalability. For the case with 512 MPI tasks, it has super-linear scalability.

\begin{figure}[!htb]
    \centering
    \includegraphics[width=0.6\linewidth, angle=270]{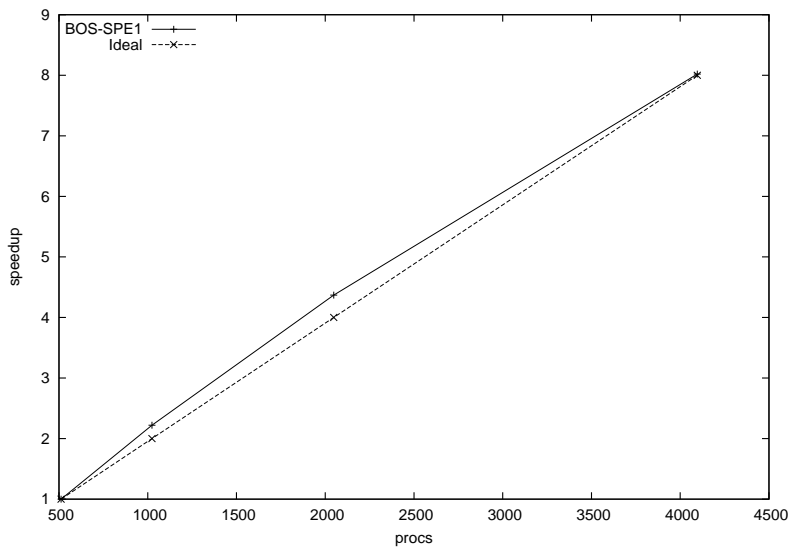}
    \caption{Scalability (speedup) of Example \ref{ex9}}
    \label{fig-ex9-cpr}
\end{figure}

\begin{example}
\label{ex9}
This case simulates a refined SPE1 model, which has 100 millions of grid cells.
The termination tolerance for inexact Newton method is $10^{-2}$ and maximal Newton iterations is 15.
The BICGSTAB linear solver is chosen and its maximal iterations is 20.
ABF decoupling strategy is enabled.
The simulation period is 10.
Summaries of numerical results are shown in Table \ref{tab-ex9-1}
and scalability curve is shown by Figure \ref{fig-ex9-cpr}.
\end{example}

\begin{table}[!htb]
\centering
\caption{Summaries of Example \ref{ex9}}
\begin{tabular}{lllllll} \hline
  \# Procs  & \# Steps & \# Newton & \# Solver & \# Avg. solver  & Time (s) & Avg. time (s)\\ \hline
512  & 27 (1) & 140 & 586 & 4.1 & 11827.9 & 84.4 \\
1024 & 27 & 129 & 377 & 2.9 & 5328.4 & 41.3 \\
2048 & 26 & 122 & 362 & 2.9 & 2708.5 & 22.2  \\
4096 & 27 & 129 & 394 & 3.0 & 1474.2 & 11.4  \\ \hline
\end{tabular}
\label{tab-ex9-1}
\end{table}

The original SPE1 project is a small one with 300 grid cells ($10 \times 10 \times 3$) and the
project is refined to 100 millions of grid cells ($1000 \times 1000 \times 100$). It has a homogeneous
geological model. Table \ref{tab-ex9-1} presents numerical summaries for nonlinear method, linear
solver and performance. All four simulations use around 27 time steps. The simulation with 512 MPI tasks
uses 140 Newton iterations, which is more than other cases due to one time step cut. The linear solver
and preconditioner are robust, which can solve a linear system in a few iterations. Again,
running time and Figure \ref{fig-ex9-cpr} show our simulator has excellent scalability.

\section{Conclusion}
Parallel reservoir simulations are studied in the paper, which are based on our in-house parallel platform. The platform
provides grids, data, linear solvers and preconditioners for reservoir simulators. A black oil
model is implemented. Numerical experiments show that our simulator has excellent scalability
and simulations can be sped up thousands of times faster.
The paper also demonstrates that parallel computing techniques are powerful tools for large-scale reservoir simulations
and IBM Blue Gene/Q system is scalable.

\section*{Acknowledgements}
The support of Department of Chemical and Petroleum Engineering, University of
Calgary and Reservoir Simulation Group is gratefully acknowledged. The research is
partly supported by NSERC/AIEE/Foundation CMG and AITF Chairs.


\begin{thebibliography}{4}

\bibitem{c1}
Coats, K., Reservoir simulation, SPE-1987-48-PEH, Society of Petroleum Engineers, 1987.

\bibitem{c2}
Coats, K.,Simulation of Steamflooding With Distillation and Solution Gas,
SPE-5015-PA, Society of Petroleum Engineers Journal, 1976, 235-247.

\bibitem{c3}
Coats, K., Reservoir Simulation: State of the Art, SPE-10020-PA, Journal of Petroleum Technology,
34(08), 1982, 1633-1642.

\bibitem{PS-Rut}
Rutledge, J., Jones, D., Chen, W., and Chung, E., The Use of Massively Parallel SIMD
    Computer for Reservoir Simulation, SPE-21213, eleventh SPE Symposium on
    Reservoir Simulation, Anaheim, 1991.

\bibitem{PS-Shi}
Shiralkar, G., Stephenson, R., Joubert, W., Lubeck, O., and van Bloemen Waanders, B.,
A production quality distributed memory reservoir simulator,
SPE Reservoir Simulation Symposium. 1997.

\bibitem{PS-Kaa}
Kaarstad, T., Froyen, J., Bjorstad, P., Espedal, M., Massively Parallel Reservoir Simulator,
SPE-29139, presented at the 1995 Symposium on Reservoir Simulation,
San Antonio, Texas, 1995.

\bibitem{PS-Kil2}
Killough, J., Camilleri, D., Darlow, B., Foster, J.,
Parallel Reservoir Simulator Based on Local
Grid Refinement, SPE-37978, SPE Reservoir
Simulation Symposium, Dallas, 1997.

\bibitem{c4}
Coats, K., A Highly Implicit Steamflood Model, SPE-6105-PA, Society of Petroleum Engineers Journal,
18(05), 1978, 369-383.

\bibitem{PS-Dogru2}
Dogru, A., Fung, L., Middya, U., Al-Shaalan, T., Pita, J.,
A next-generation parallel reservoir simulator for giant reservoirs,
SPE/EAGE Reservoir Characterization \& Simulation Conference. 2009.

\bibitem{phg}
{Zhang, L.,}, {A Parallel Algorithm for Adaptive Local Refinement of Tetrahedral Meshes
Using Bisection}, Numer. Math.: Theory, Methods and Applications, 2009, 2, 65--89.

\bibitem{c6}
Coats, K., An Equation of State Compositional Model, SPE-8284-PA, Society of Petroleum
Engineers Journal, 20(05), 1980, 363-376.

\bibitem{phg-quad}
{Zhang, L., Cui, T., and Liu, H.,}, {A set of symmetric quadrature rules on triangles and tetrahedra},
J. Comput. Math, 2009, 27(1), 89--96.

\bibitem{CPR-old}
Wallis, J., Kendall, R., and Little, T.,
Constrained residual acceleration of conjugate residual methods,
SPE Reservoir Simulation Symposium, 1985.

\bibitem{CPR-cao}
Cao, H., Schlumberger, T., Hamdi, A., Wallis, J., Yardumian, H.,
Parallel scalable unstructured CPR-type linear solver for reservoir simulation.
SPE Annual Technical Conference and Exhibition. 2005.

\bibitem{Study-Two-Stage}
Al-Shaalan, T., Klie, H., Dogru, A., Wheeler, M.,
Studies of Robust Two Stage Preconditioners for the Solution of Fully Implicit Multiphase Flow Problems.
SPE Reservoir Simulation Symposium. 2009.

\bibitem{FASP}
Hu, X., Liu, W., Qin, G., Xu, J., Zhang, Z.,
Development of a fast auxiliary subspace pre-conditioner for numerical reservoir simulators,
SPE Reservoir Characterisation and Simulation Conference and Exhibition. 2011.

\bibitem{c5}
Coats, K., Effects of Grid Type and Difference Scheme on Pattern Steamflood Simulation Results,
SPE-11079-PA, Journal of Petroleum Technology, 38(05), 1986, 557-569.


\bibitem{FASP2}
Feng, C., Shu, S., Xu, J., and Zhang, C.,
A Multi-Stage Preconditioner for the Black Oil Model and Its OpenMP Implementation,
21st International Conference on Domain Decomposition Methods, 2012, France.

\bibitem{Book-Chen}
Chen, Z., Huan, G., and Ma, Y., Computational methods for multiphase flows in porous media, Vol. 2. Siam, 2006.

\bibitem{HYPRE2}
Falgout, R., and Yang, U., HYPRE: A library of high performance preconditioners,
Lecture Notes in Computer Science, Springer Berlin Heidelberg, 2002. 632-641.

\bibitem{bos-pc}
Liu, H., Wang, K., and Chen, Z., A family of constrained pressure residual preconditioners for parallel reservoir
simulations, Numerical Linear Algebra with Applications, DOI: 10.1002/nla.2017.

\bibitem{SPE10}
Christie, M., and Blunt, M.,
Tenth SPE comparative solution project: A comparison of upscaling techniques.
SPE Reservoir Evaluation \& Engineering 4.4 (2001): 308-317.

\bibitem{mlp}
Baohua Wang, Shuhong Wu, Qiaoyun Li, Xiaobo Li, Hua Li, Chensong Zhang, Jinchao Xu,
A Multilevel Preconditioner and Its Shared Memory Implementation for New Generation Reservoir Simulator,
SPE-172988-MS,
SPE Large Scale Computing and Big Data Challenges in Reservoir Simulation Conference and Exhibition, 15-17 September, Istanbul, Turkey, 2014.

\bibitem{kwang}
Wang, K., Zhang, L., and Chen, Z., Development of Discontinuous Galerkin Methods and a Parallel Simulator for Reservoir Simulation,
SPE-176168-MS, SPE/IATMI Asia Pacific Oil \& Gas Conference and Exhibition, 20-22 October, Nusa Dua, Bali, Indonesia, 2015.

\bibitem{PWM}
Peaceman D., Interpretation of Well-Block Pressures in Numerical Reservoir Simulation,
SPE-6893, 52nd Annual Fall Technical Conference and Exhibition, Denver, 1977.

\end{thebibliography}
\end{document}